\documentclass[a4paper,10pt,twoside]{article}
\usepackage[INRIA]{lipinria}
\usepackage{figlatex}
\usepackage{pdfswitch}
\usepackage[latin1]{inputenc}
\usepackage{subfig}
\usepackage{url} 
\usepackage{xspace}

\newcommand{\app}[1]{\textsc{#1}}
\newcommand{\nws}{NWS\xspace}
\def\env{\textsc{ENV}\xspace}
\def\alnem{\textsc{ALNeM}\xspace}
\def\nws{\textsc{NWS}\xspace}
\def\gras{\textsc{GRAS}\xspace}
\def\simgrid{\textsc{SimGrid}\xspace}
\def\nsl{\textsc{NetSolve}\xspace}

\def\clique{\textsc{Clique}\xspace}
\def\treelat{\textsc{TreeLat}\xspace}
\def\treebw{\textsc{TreeBW}\xspace}
\def\aggregate{\textsc{Aggregate}\xspace}
\def\imptreelat{\textsc{ImpTreeLat}\xspace}
\def\imptreebw{\textsc{ImpTreeBW}\xspace}
\def\improving{\textsc{Improving}\xspace}

\def\eg{e.g.,\xspace}
\def\ie{i.e.,\xspace}

\usepackage{manfnt}

\newcommand{\FFIXME}[1]{%
}
\usepackage{RR}

\begin{document}
\RRNo{6133}
\RRInumber{2007-08}
\RRIdate{February 2007}

\RRItitle{A First Step Towards\\ Automatically Building Network Representations} 
\RRItitre{Une première étape vers la construction automatique de modèles de réseau d'interconnexion}
\RRIthead{A First Step Towards Automatically Building Network Representations}     

\RRIauthor{Lionel Eyraud-Dubois\and Arnaud Legrand\and Martin Quinson\and Fr\'{e}d\'{e}ric Vivien}
\RRIahead{L. Eyraud-Dubois\and A. Legrand\and M. Quinson\and F. Vivien}

\RRIkeywords{Network model, topology reconstruction, Grids}
\RRImotscles{Modélisation de réseau, reconstruction de topologies, grilles}

\RRIabstract{To fully harness Grids, users or middlewares must have
  some knowledge on the topology of the platform interconnection
  network. As such knowledge is usually not available, one must uses
  tools which automatically build a topological network model through
  some measurements. In this article, we define a methodology to
  assess the quality of these network model building tools, and we
  apply this methodology to representatives of the main classes of
  model builders and to two new algorithms. We show that none of the
  main existing techniques build models that enable to accurately
  predict the running time of simple application kernels for actual
  platforms. However some of the new algorithms we propose give
  excellent results in a wide range of situations.}

\RRIresume{Afin de tirer le meilleur parti des grilles, les
  utilisateurs et les intergiciels doivent avoir connaissance de la
  topologie du réseau d'interconnexion de la plate-forme utilisée.
  Comme cette connaissance n'est généralement pas disponible \emph{a
    priori}, on doit avoir recours à des outils construisant un modèle
  du réseau d'interconnexion à partir de mesures. Dans cet article
  nous définissons une méthodologie pour évaluer la qualité de ces
  outils de construction de modèles de réseau, et nous l'appliquons à
  des représentants des principaux types de reconstructeurs de
  topologies, ainsi qu'à deux nouveaux algorithmes. Nous montrons
  qu'aucune des techniques existantes ne produit des modèles qui
  permettent de prédire avec précision le temps d'exécution sur les
  plates-formes actuelles de simples noyaux d'applications. Au
  contraire, un des nouveaux algorithmes obtient de très bons
  résultats dans des situations très variées.}

\RRItheme{\THNum}
\RRIprojets{ALGORILLE, GRAAL, MESCAL}

\RRImaketitle

\noindent

\section{Introduction}

Grids are parallel and distributed systems that result from the
sharing and aggregation of resources distributed between several
geographically distant organizations~\cite{Fosterbook2004}. Unlike classical
parallel machines, Grids present heterogeneous and sometimes even
non-dedicated capacities. Gathering accurate and relevant information
about them is then a challenging issue, but it is also a necessity.
Indeed, the efficient use of Grid resources can only be achieved
through the use of accurate network information. Qualitative
information such as the network topology is crucial to achieve tasks
such as running network-aware applications~\cite{LegrandReRoVi04},
efficiently placing servers~\cite{ArticleChouhan.CDCV_IJHPCA06}, or predicting and optimizing
collective communications performance~\cite{kielmann99magpie}.

However, the description of the structure and characteristics of the
network interconnecting the different Grid resources is usually not
available to users. This is mainly due to security (fear of Deny Of
Service attacks) and privacy reasons (ISP do not want you to know
where their bottlenecks are). Hence a need for tools which
automatically construct models of platform networks. Many tools and
projects provide some network information. Some rely on simple ideas
while others use very sophisticated measurement techniques. Some of
these techniques, though, are sometimes ineffective in Grid
environments due to security issues.  Anyway, to the best of our
knowledge, these different techniques have never been compared
rigorously in the context of Grid computing platforms. Our aim is to
define a methodology to assess the quality of network model building
tools, to apply it to representatives of the main classes of model
builders, to identify weaknesses of existing approaches, and to
propose new model building algorithms.

The main contributions of this paper are the definition of a
methodology to assess the quality of reconstruction algorithms, the
design of two new reconstruction algorithms, and some evaluations that
highlight the weaknesses of classical algorithms and demonstrate the
superiority of one of our new algorithms.

The rest of this article is organized as follows.  In
Section~\ref{sec:related}, we review the main observation techniques
and we identify some that are effective in Grid environments. In
Section~\ref{sec:methods} we review existing reconstruction algorithms
and we identify a few representative ones. Based on the analysis of
potential weaknesses of these algorithms, we propose two new
algorithms. In Section~\ref{sec:metrics} we present our methodology to
assess the quality of reconstruction algorithms. In
Section~\ref{sec:synthetic}, we evaluate through simulation the
quality of the studied reconstruction algorithms with respect to the
proposed metrics. This evaluation is performed on models of real
platforms and on synthetic models.

\section{Related Work}
\label{sec:related}

Network discovery tools have received a lot of attention in recent
years. However, most of them are not suited to Grids.  Indeed, much of
the previous work (\eg Remos~\cite{remos,miller00remos}) rely on
low-level network protocols like SNMP or BGP, whose usage is generally
restricted for security reasons (it is indeed possible to conduct Deny
Of Service attacks by flooding the routers with requests).

As a matter of fact, in a Grid environment, even traceroute or
ping-based tools (\eg TopoMon~\cite{TopoMon}, Lumeta~\cite{lumeta},
IDmaps~\cite{al00idmaps}, Global Network
Positioning~\cite{ng01predicting}) are getting less and less
effective. Indeed, these tools rely on ICMP which is more and more often
disabled by administrators, once again to avoid Deny Of Service attacks
based on flooding.  For example, the \textbf{Skitter}
project~\cite{caida}, which keeps track of the evolution of
the macroscopic connectivity and performance of the Internet, reports
that in 5 years of measurements the number of hosts replying to ICMP
requests decreases by 2 to 3\% per month.

Even if recent works have proposed similar or even better
functionalities without relying on ICMP, some of them (\eg
pathchar~\cite{downey99pathchar}) require specific privileges, which
make them unusable in our context. It is mandatory to rely on tools
that only use \emph{application-level measurements}, \ie measurements
that can be done by any application running on a computing Grid
without any specific privilege.
This comprises the common end-to-end measurements, like bandwidth and
latency, but also interference measurements (\ie whether a
communication between two machines $A$ et $B$ has non negligible
impact on the communications between two machines $C$ et $D$). Many
projects rely on this type of measurements.

An example is the \nws (Network Weather Service)~\cite{nws} software,
which constitutes a \textit{de facto} standard in the Grid community
as it is used by major \textit{Grid middlewares} like
Globus~\cite{foster97globus} or \textit{Problem Solving Environments}
(PSEs) like DIET~\cite{ArticleCaron.CD_IJHPCA06},
\nsl~\cite{CasDon97}, or NINF~\cite{ninfg} to gather information about
the current state of a platform and to predict its evolutions.  NWS is
able to report the end-to-end bandwidth, latency, and connection time,
which are typical application-level measurements.  However, the NWS
project focuses on quantitative information and does not provide any
kind of topological information.  It is however natural to address
this issue by aggregating all NWS information in a single clique graph
and use this labeled graph as a network model.

In another example, interference measurements have been used in
\env~\cite{env} and enabled to detect, to some extent, whether some
machines are connected by a switch or a hub.

A last example is ECO~\cite{eco-lowekamp}, a collective communication
library, that uses plain bandwidth and latency measurements to propose
optimized collective communications (\eg broadcast, reduce,
etc.). These approaches have proved to be very effective in practice,
but they are generally very specific to a single problem and we are
looking for a general approach.

\section{Studied Reconstruction Algorithms}
\label{sec:methods}

We are thus looking for a tool based on application-level measurements
that would enable any network-aware application to benefit from
reasonably accurate information on the network topology. In most
previous works, the underlying network topology is either a
clique~\cite{nws,eco-lowekamp} or a tree~\cite{mint,env}. Our
reference reconstruction algorithms are thus clique, minimal spanning
tree on latencies, and maximal spanning tree on bandwidths.
As our experiments show (Section~\ref{sec:synthetic}), these methods
produce very simple graphs, and often fail to provide a realistic view
of platforms. We thus designed two new reconstruction algorithms, as a
first step towards a better reconstruction. The first algorithm aims
at improving an already built topology and is meant to be used to
improve an existing spanning tree. The second one reconstructs a
platform model from scratch, by growing a set of connected nodes.
Both algorithms keep track of the routing while building their model,
to be able to correct a route connecting two nodes whose latency was
previously inaccurately predicted. We focus on latency rather than on
bandwidth as bandwidths are less discriminant.

\subsection{Algorithm \textnormal{\improving}}

Algorithm \improving is based on the observation that if the latency
between two nodes is badly over-predicted by the current route
connecting them, an extra edge should be inserted to connect them
through an alternate and more accurate route. Among all pairs of
``badly connected'' nodes, we pick the two nodes with the smallest
possible measured latency, and we add a direct edge between them.
Each time \improving adds an edge, for each pair of nodes whose
latency is over-predicted, we check whether that pair cannot be better
connected through the just introduced edge, and we update the routing
if needed. This edge addition procedure is repeated until all
predictions are considered sufficiently accurate.  The accuracy of
predictions is necessarily arbitrary. In our implementation, it
corresponds to a deviation of less than 10\% from actual measurements.

\subsection{Algorithm \textnormal{\aggregate}}

Algorithm \aggregate uses a more local view of the platform.
It expands a set of already connected nodes, starting with the two
closest nodes   in terms of latency. At each step, \aggregate
connects a new \emph{selected} node to the already connected ones.
The selected node is the one closest to the connected set in terms of
latency. \aggregate iteratively adds edges so that each route
from the selected node to a connected node is sufficiently accurate.
Added edges are greedily chosen starting from the edge yielding a
sufficiently accurate prediction for the largest number of routes from
the selected node to a connected node.
We slightly modified this scheme to avoid adding edges that will later
become redundant. A new edge is added only if its latency is not
significantly larger (meaning less than 50\% larger) than that of the
first edge added to connect the selected node. Because of this
change, we may move to a new selected node while not all the
routes of the previous one are considered accurate enough.  We thus
keep a list of \emph{inaccurate} routes. For each edge addition we
check whether the new edge defines a route rendering accurate an
inaccurate route. When all nodes are connected, we add edges to
correct all remaining inaccurate routes, starting with the route of
lowest latency.

\section{Assessing the Quality of Reconstructions}
\label{sec:metrics}

We want to thoroughly assess the quality of reconstruction algorithms.
To fairly compare various topology mapping algorithms, we have
developed \alnem (Application Level Network Mapper).  \alnem is
developed with \gras~\cite{gras} that provides a complete API to
implement distributed application on top of heterogeneous platforms.
Thanks to two different implementations of \gras, \alnem can work
seamlessly on real platforms as well as on simulated platforms with
\simgrid~\cite{simgrid}. \alnem is made of three main parts:
\begin{enumerate}
\item a measurement repository (MySQL database);
\item a distributed collection of sensors performing bandwidth,
  latency, and interference measurements;
\item a topology builder implementing reconstruction algorithms which
  use the repository.
\end{enumerate}

The evaluation of the quality of model builders is not an easy task.
To perform such an evaluation, we use three different and
complementary approaches. For each approach, we will consider a series
of original platforms; and for each of these platforms we will compare
the original platform and the models built from it.

The three approaches can be seen as different point of views on the
models: a structural one, a communication-level one, and an
application-level one.

To assess the quality of model builders, we use two different and
complementary approaches. For both approaches, we consider a series of
original platforms; and for each platform we compare the original
platform and the models built from it. The two approaches can be seen
as different points of view on models: a communication-level one and
an application-level one.

\subsection{End-to-End Metric}

A platform model is ``good'' if it allows to accurately predict the
running time of applications. The accuracy of the prediction depends
on the model capacity to render different aspects and characteristics
of the network. Most of the time, researchers only focus on bandwidth
predictions. However, latencies and interferences can also greatly
impact an application performance. Therefore, we consider the three
following characteristics:

\subsubsection{Bandwidth} This is the most obvious characteristic. We
need to know the bandwidth available between processors as soon as the
different tasks of an application, or the different applications run
concurrently, send messages of different lengths.

\subsubsection{Latencies} Obviously, latencies are very important for
small messages. They are, however, often overlooked in the context of
Grid computing, because of the usual assumption that in this framework
processes only exchange large messages. Casanova presented an
example~\cite{Casanova2004} on the TeraGrid platform where one third
of the time needed to transfer a 1 GByte of data would be due to
latencies. Therefore, latencies cannot always be neglected even for
large messages, and models must be able to predict them accurately. In
practice, latencies can range from 0.1~ms for intra-cluster
communications, to more than 300~ms for intercontinental satellite
communications. Applications must be aware of the magnitude of the
latencies to be able to organize their communications
efficiently.

\subsubsection{Interferences} Many distributed applications use
collective communications (\eg broadcasts or all-to-all) or, more
generally, independent communications between disjoint pairs of
processors. The only knowledge of the available latencies and
bandwidths between any two pairs of processors does not allow to
predict the time needed to realize two communications between two
disjoint pairs of processors.  Indeed, this depends on whether the two
communications use a same physical link\footnote{In some cases, two
  communications sharing the same physical communication link do not
  interfere with each other.  This may happen, for example, when the
  only shared communication links are backbones, as exemplified by
  Casanova~\cite{Casanova2004}.}.  Legrand, Renard, Robert, and Vivien
have shown~\cite{LegrandReRoVi04} that knowing the network topology,
and thus being able to predict communication interferences, enable to
derive algorithms far more efficient in practice.

\vspace{\baselineskip}

\subsubsection*{Methodology} Our evaluation methodology is based on
simulations. Given one original platform, we measure the end-to-end
latencies and bandwidths between any two processors. We also measure
the end-to-end bandwidths obtained when any two pairs of processors
simultaneously communicate. We then perform the same measurement on
the reconstructed models. To compare the results, we build an accuracy
index for each reconstruction algorithm, each graph, and each studied
network characteristic. For latencies and bandwidths,
following~\cite{TsafrirEtFe06}, we define \emph{accuracy} as the
maximum of the two ratios $x_R/x_M$ and $x_M/x_R$, where $x_R$ is the
reconstructed value and $x_M$ is the original measured one. We compute
the accuracy of each pair of nodes, and then the geometric mean of all
accuracies.

\subsection{Application-Level Measurements}

To simultaneously analyze a combination of the characteristics studied
with end-to-end measurements, we also compare, through simulations,
the performance of several classical distributed routines when run on
the original graph and on each of the reconstructed ones. This allows
us to evaluate the predictive power of the reconstruction algorithms
with applications with more complex but realistic communication
patterns. This approach gives us an evaluation of the quality of
reconstructions at the application level, rather than at a single
communication level like end-to-end measurements.

We study the following simple distributed algorithms (listed from the
simplest communication pattern to the most complicated one):
\begin{itemize}
\item \textbf{Token ring:} a token circulates three times along a
  randomly built ring (the ring structure has a priori no correlation
  with the interconnection network structure).
\item \textbf{Broadcast:} a randomly picked node sequentially sends
  the same message to all the other nodes.
\item \textbf{All-to-all:} all the nodes simultaneously perform a
  broadcast.
\item \textbf{Parallel matrix multiplication:} a matrix multiplication is
  realized using ScaLAPACK outer product algorithm~\cite{Scalapack97}.
\end{itemize}

This evaluation must be done through simulations. Indeed, the
measurements on the reconstructed models can obviously not be done
experimentally. Furthermore, the comparison of experimental (original
platform) and simulated (reconstructed models) measurements would
introduce a serious bias in the evaluation framework, due to the
differences between the actual world and the simulator.


\section{Experimental Results}
\label{sec:synthetic}

We present two types of experiments: the first one is based on a
modeling of a real network architecture, while for the second one we
generated synthetic platforms using GridG~\cite{GridG}.  As stated in
section~\ref{sec:methods}, we evaluate several reconstruction
algorithms. In addition to our three reference reconstruction methods
(\clique, minimal spanning tree on latencies (\treelat), and maximal
spanning tree on bandwidths(\treebw)), we analyze the performance of
the \aggregate algorithm, and of the \improving procedure applied to
both spanning trees: \imptreelat and \imptreebw.

\subsection{Renater}

Renater\footnote{\url{http://www.renater.fr}} is the French public
network infrastructure that connects major universities. Thanks to a
collection of accounts in several universities, we were able to
measure latencies and bandwidths between the corresponding hosts. For
security reasons, these measurements were performed using the most
basic tools, namely \texttt{ping} for latency and \texttt{scp} of
bandwidths. Thanks to the topology information available on the
Renater website we created a model of this network, that we annotated
with the bandwidths and latencies we measured. We then executed our
reconstruction algorithms on the obtained model. 

\FFIXME{Pour info, ce qui suit en commentaire, c'est hors simulation. 
  We have also compared the predictions on end-to-end performance to
  the real measurements; the results are shown on
  figure~\ref{fig:renater_rl_meas}. Commentaire : les bandes passantes
  mesurees sont tres asymetriques, et l'on n'arrive pas a bien les
  representer, meme avec Clique. Par contre, pour les latences, on
  arrive a donner des informations interessantes. Au passage, les
  arbres couvrant ont 10 arêtes, la clique 110, le graphe d'ImpTreeLat
  en a 16, et ceux d'ImpTreeBW et d'Aggregate en ont 19 ; on peut dire
  qu'on arrive a bien predire les valeurs de latence avec peu d'aretes
  : on trouve bien la structure du graphe.  
  peut enlever la partie RL. LED.  \begin{figure}[htb] \centering
  \includegraphics[width=0.3\linewidth]{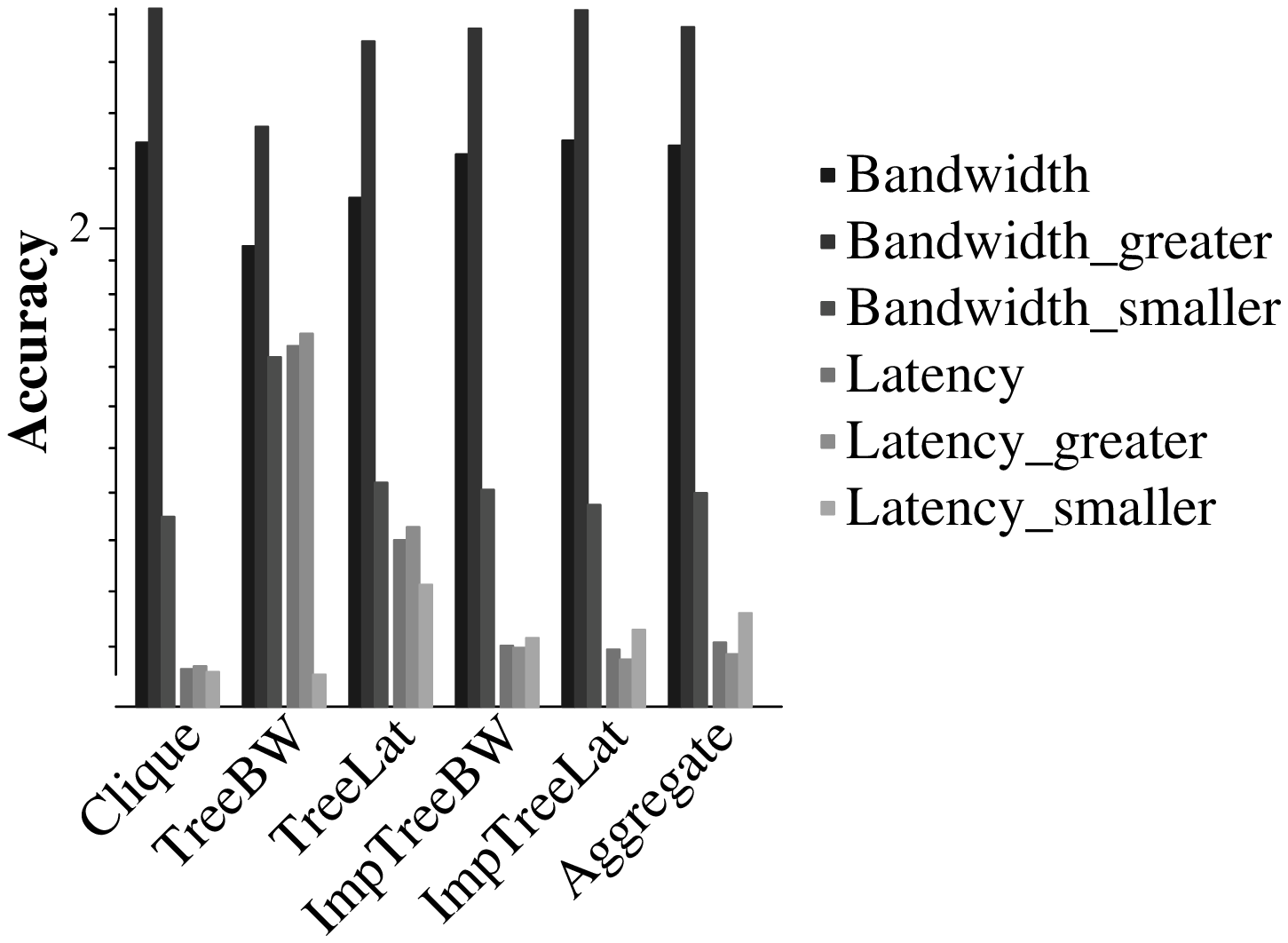}%
  \caption{Results from the real measurements on the Renater
  platform.}  \label{fig:renater_rl_meas} \end{figure} }

%
%

\begin{figure}[htb]
\centering
\hspace{2pt}
\subfloat[End-to-end metrics]{\label{fig:renater_simul_meas}%
\includegraphics[scale=0.42]{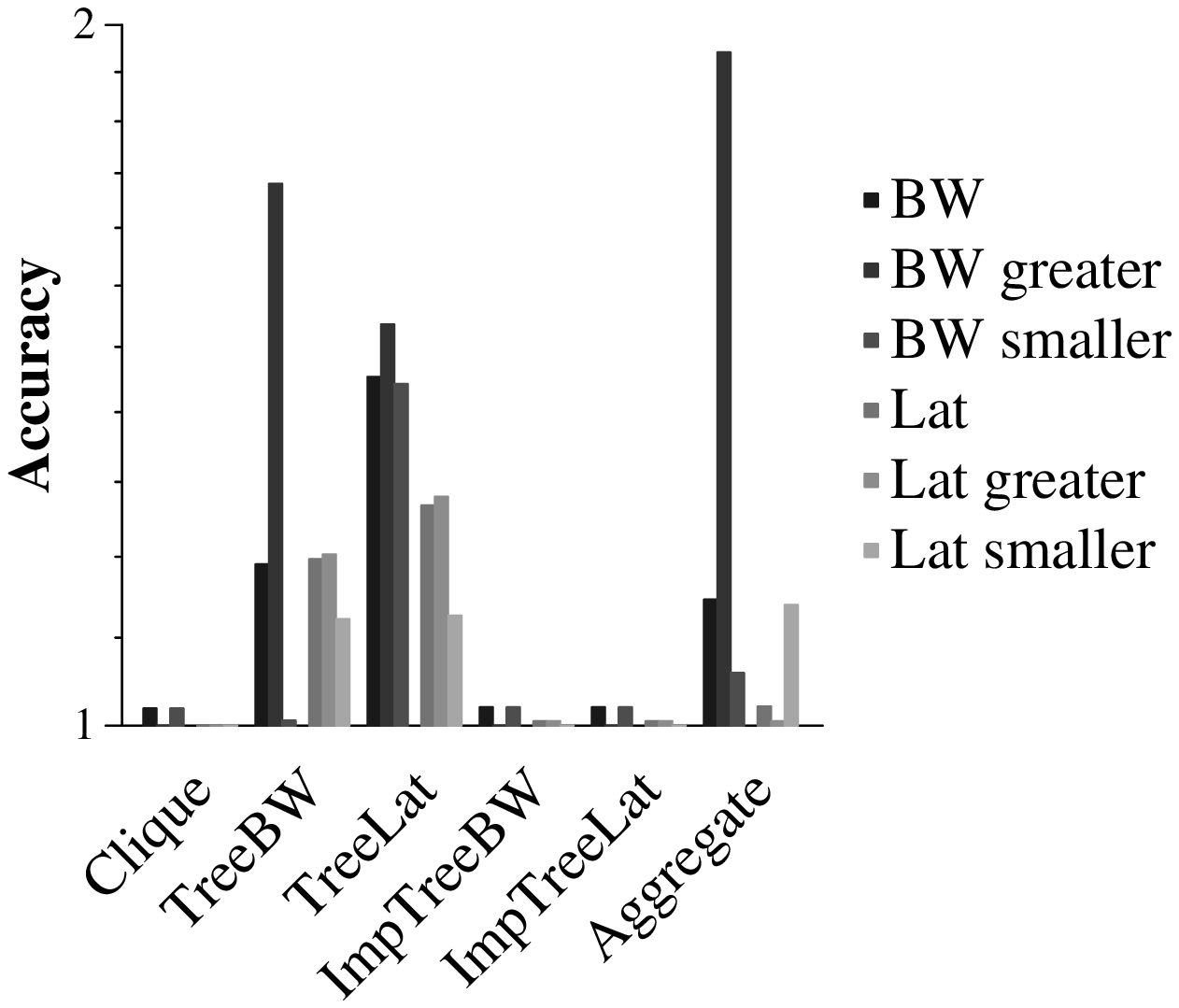}}
\hspace{2pt}
\subfloat[Applicative metrics]{\label{fig:renater_simul_app}%
\includegraphics[scale=0.42]{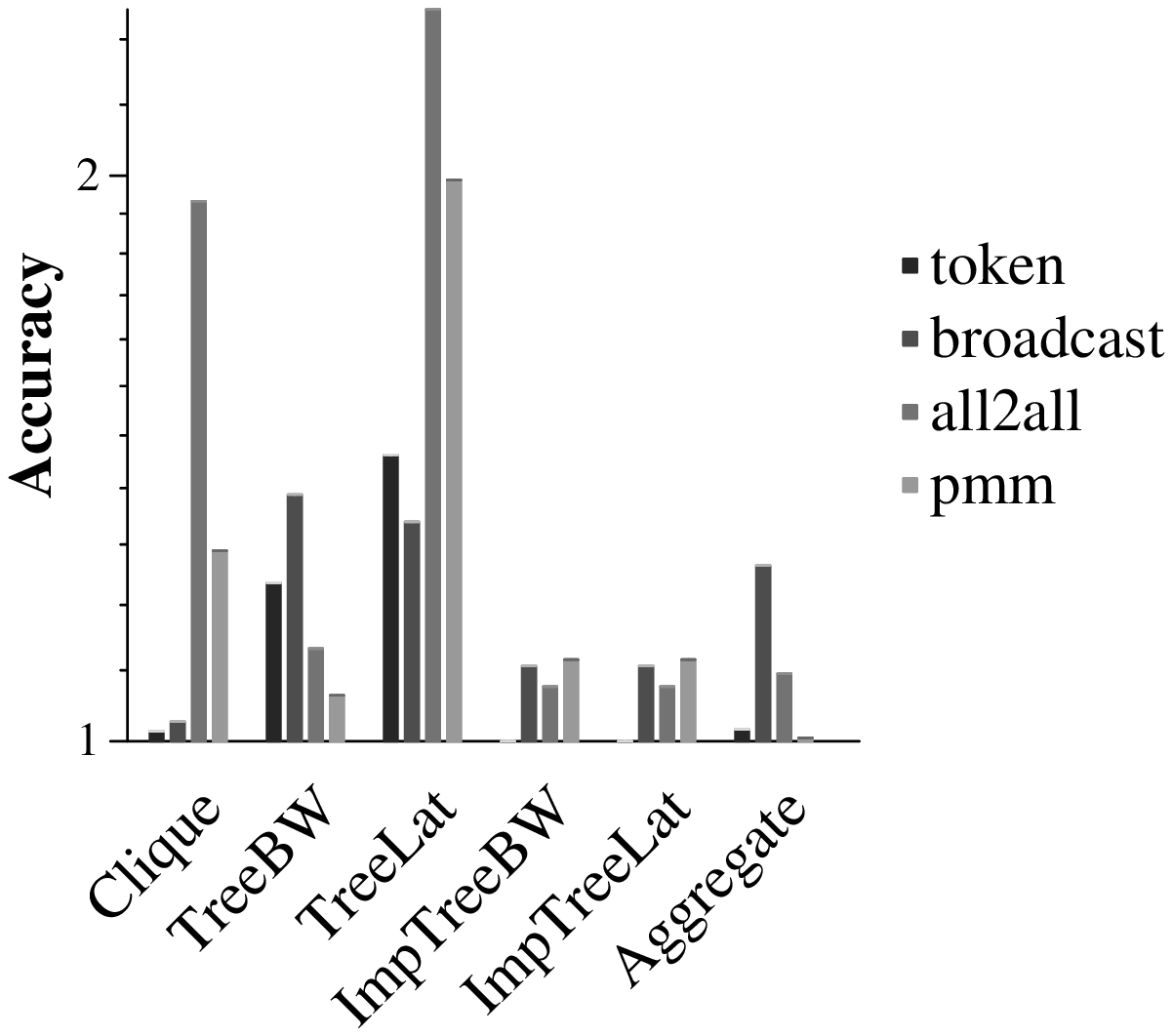}}
\hspace{2pt}

\subfloat[Interferences]{\label{fig:renater_simul_interf}%
\includegraphics[scale=0.42]{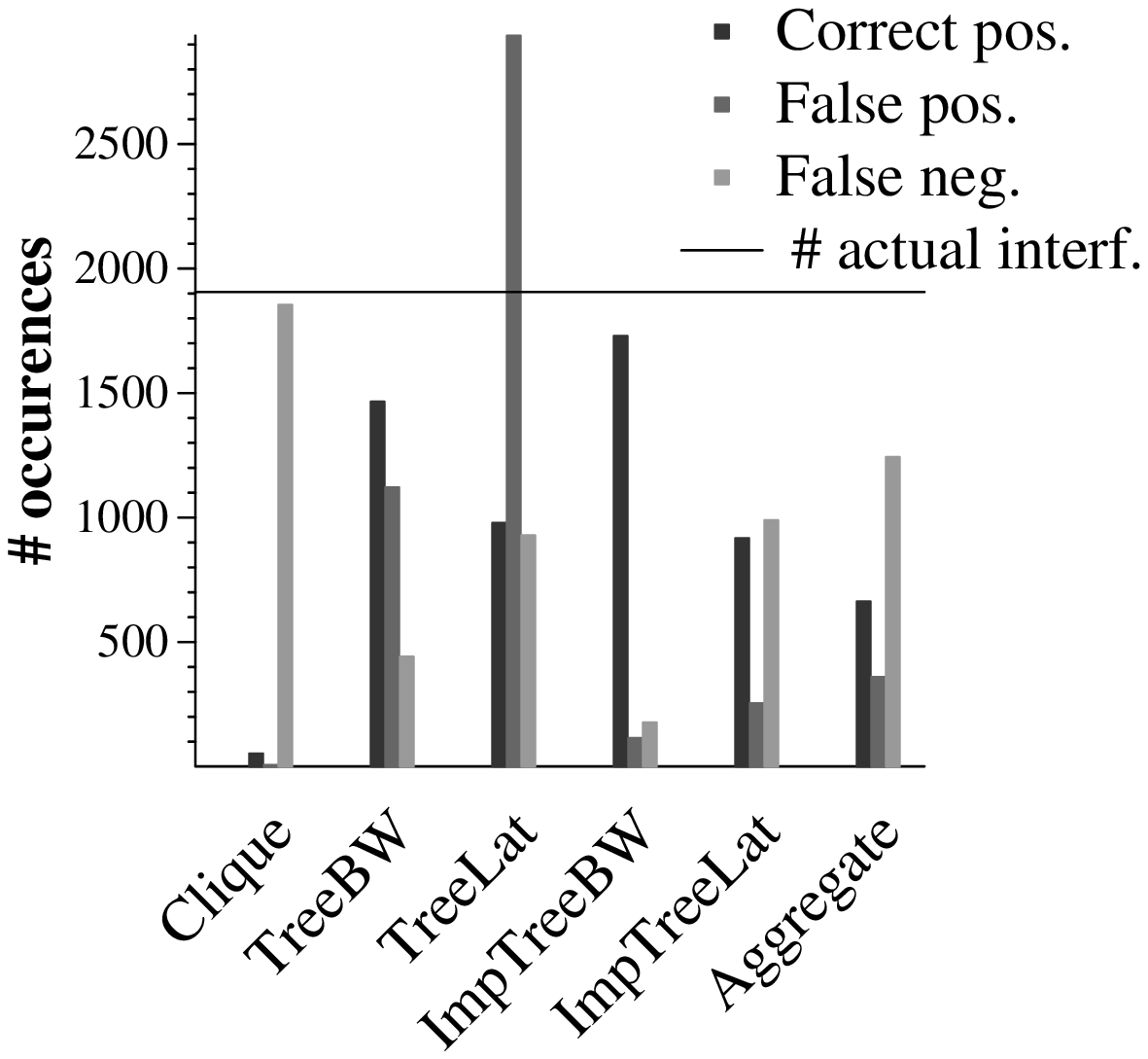}}
\caption{Simulated tests on the Renater platform.}
\label{fig:renater_tests}
\end{figure}

Figure~\ref{fig:renater_tests} shows the evaluation of the
reconstructed topologies. For end-to-end metrics, we plotted the
average accuracy for both latency and bandwidth, and we  also
detailed the average accuracy for over- and under-predicted values.
Unsurprisingly, \clique has excellent end-to-end performances whereas
\treelat and \treebw have poor ones. \aggregate over-estimates
bandwidth for a few couples, but both \imptreelat and
\imptreebw have excellent end-to-end performances.  

Regarding applicative performance, \clique is unsurprisingly good for
\app{token} and \app{broadcast} where there is always at most one
communication at a time and very bad for \app{all2all} and \app{pmm}.
\imptreelat and \imptreebw are once again equivalent and now clearly
have much better results than any other heuristics. They are actually
within 10\% of the optimal solution for all applicative performances.
Last, the interference evaluation
(Figure~\ref{fig:renater_simul_interf}) enables us to distinguish
\imptreebw and \imptreelat. \imptreebw accurately predicts more than
95\% of interferences whereas \imptreelat overestimates 50\% of
interferences!

This experiment shows that our reconstruction algorithms are able to
yield platforms with good predictive power. It also suggests that our
\imptreebw algorithm can provide very good reconstructions. The
good performance of \imptreebw may be explained by the fact that this
is the only algorithm which builds a non trivial graph (\ie, not a
clique) while using both the information on latencies and bandwidths.
However, these encouraging results obtained on a realistic platform
must be confirmed by a more comprehensive set of experiments, using a
large number of different platforms, which we do in the next section.



\subsection{GridG}

For a thorough validation of our algorithms, we used the GridG
platform generator~\cite{GridG} to study realistic
Internet-like platforms, which may be different from the very few
platforms we can access and thus test directly. In this experiment, we
generated 2 different kinds of platforms: in the first group, all of
the hosts are known to the measurement procedure, which means that it
is possible to deploy a process on all internal routers of the
platform. In the second group, only the external hosts are known to
the algorithms. For each group, we generated 40 different platforms,
each of them containing about 60 hosts.

\begin{figure*}
\centering
\subfloat[End-to-end metrics]{\label{fig:gridg_meas_without}%
\includegraphics[scale=0.42]{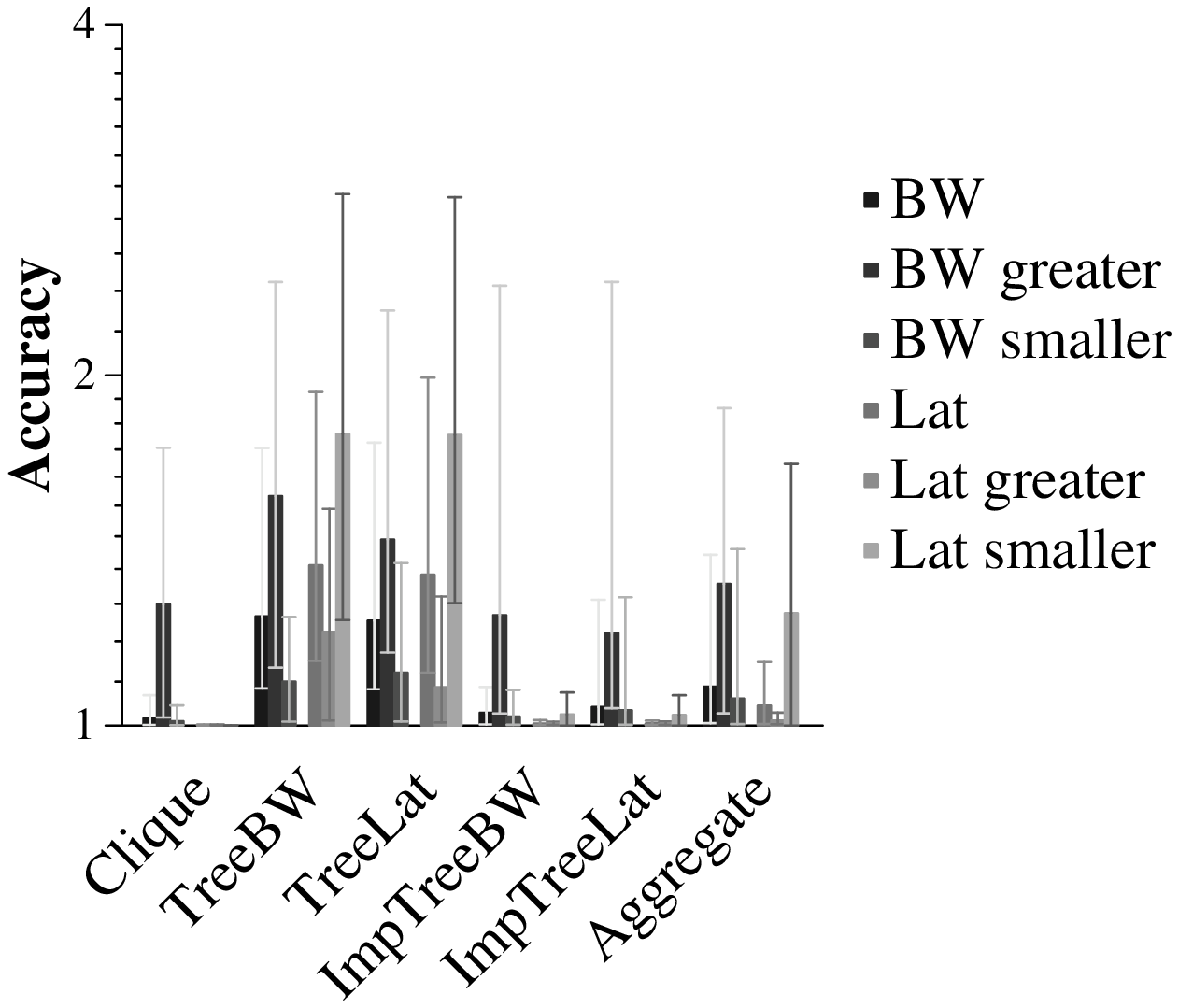}}%
\hspace{0.1\linewidth}
\subfloat[Applicative metrics]{\label{fig:gridg_app_without}%
\includegraphics[scale=0.42]{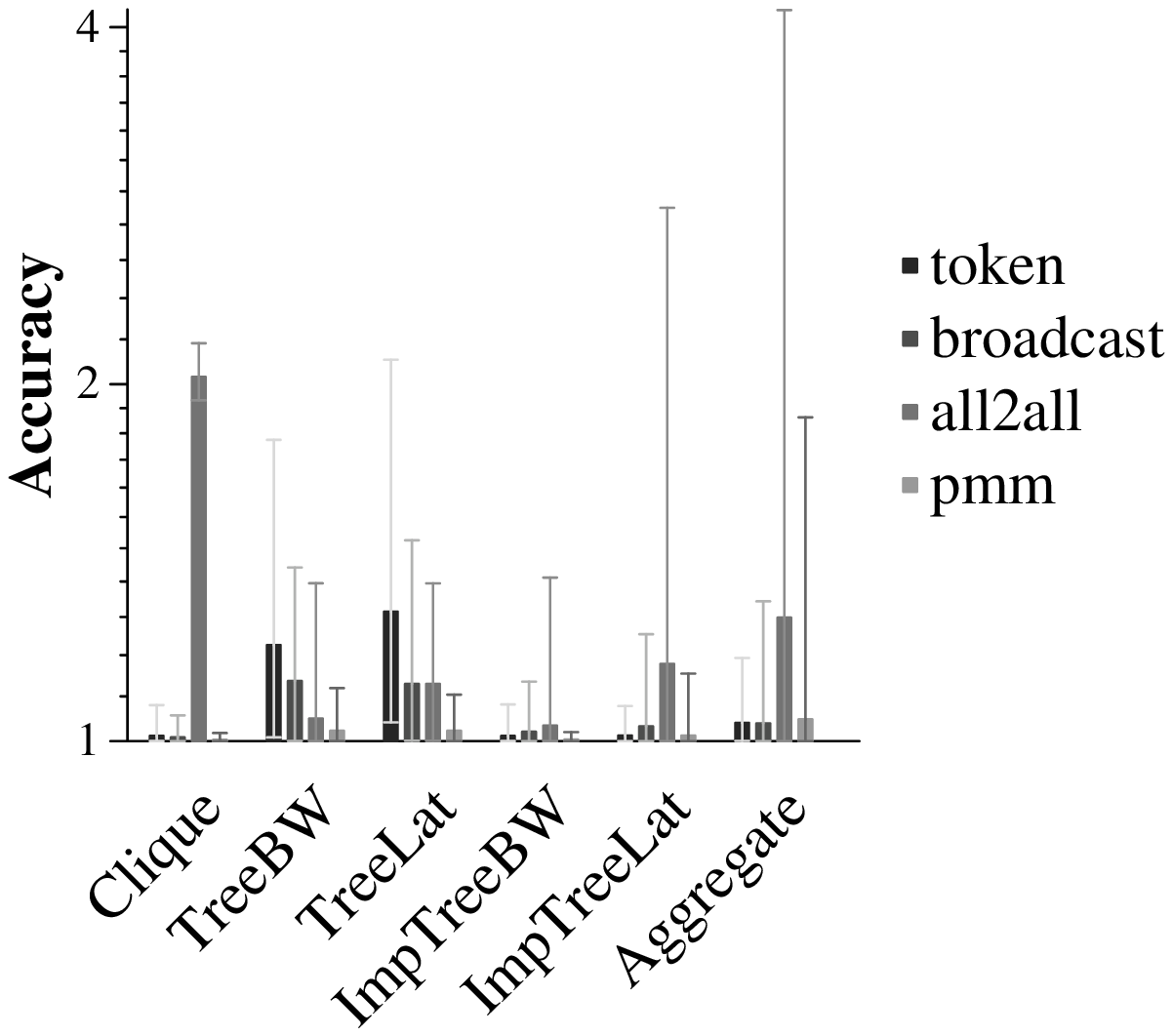}}
\caption{Simulated tests on the GridG platforms, with processes on
  every host.}
\label{fig:gridg_tests_without_routers}
\end{figure*}

\begin{figure*}
\centering
\subfloat[End-to-end metrics]{\label{fig:gridg_meas_with}%
\includegraphics[scale=0.42]{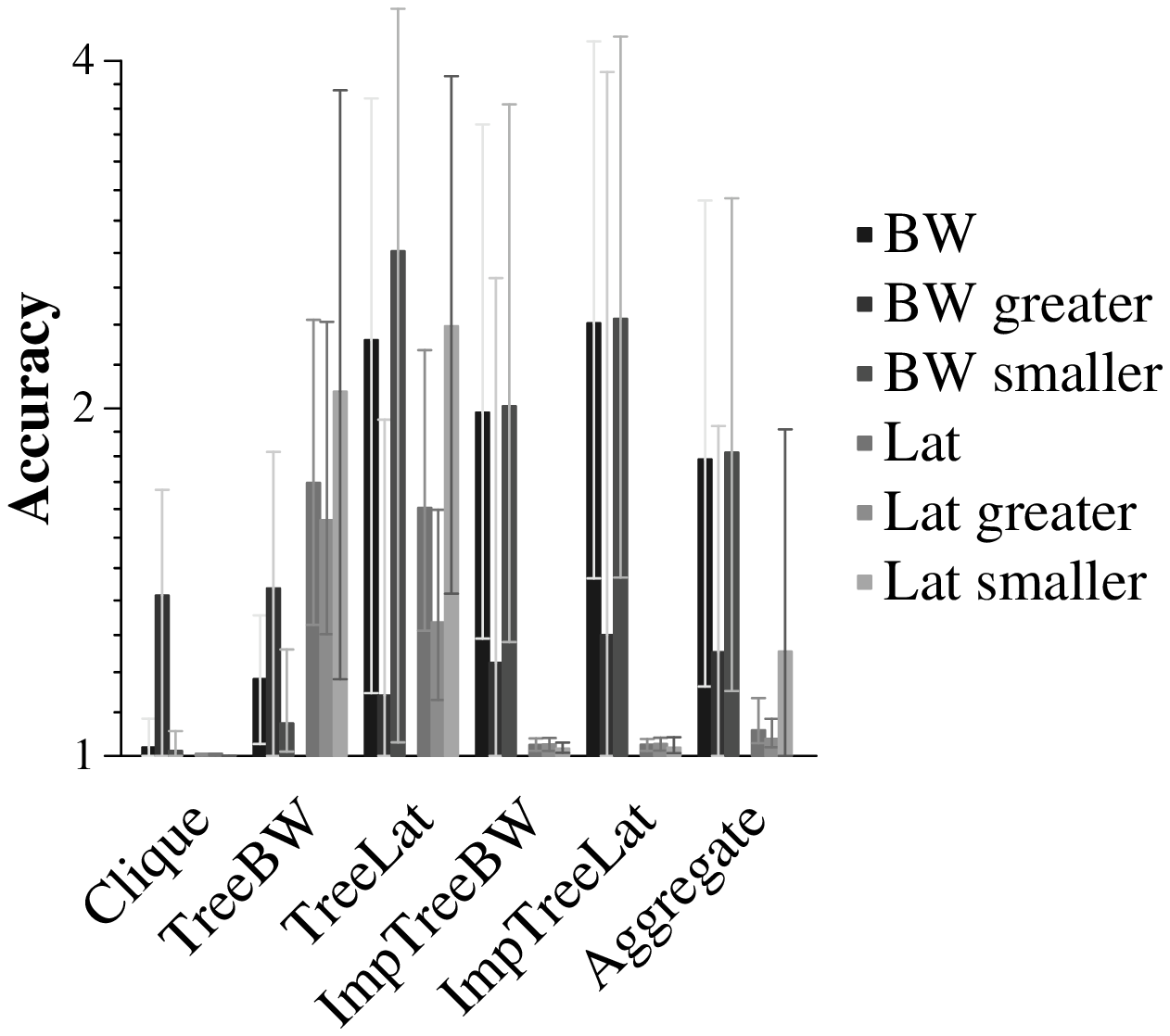}}%
\hspace{0.1\linewidth}
\subfloat[Applicative metrics]{\label{fig:gridg_app_with}%
\includegraphics[scale=0.42]{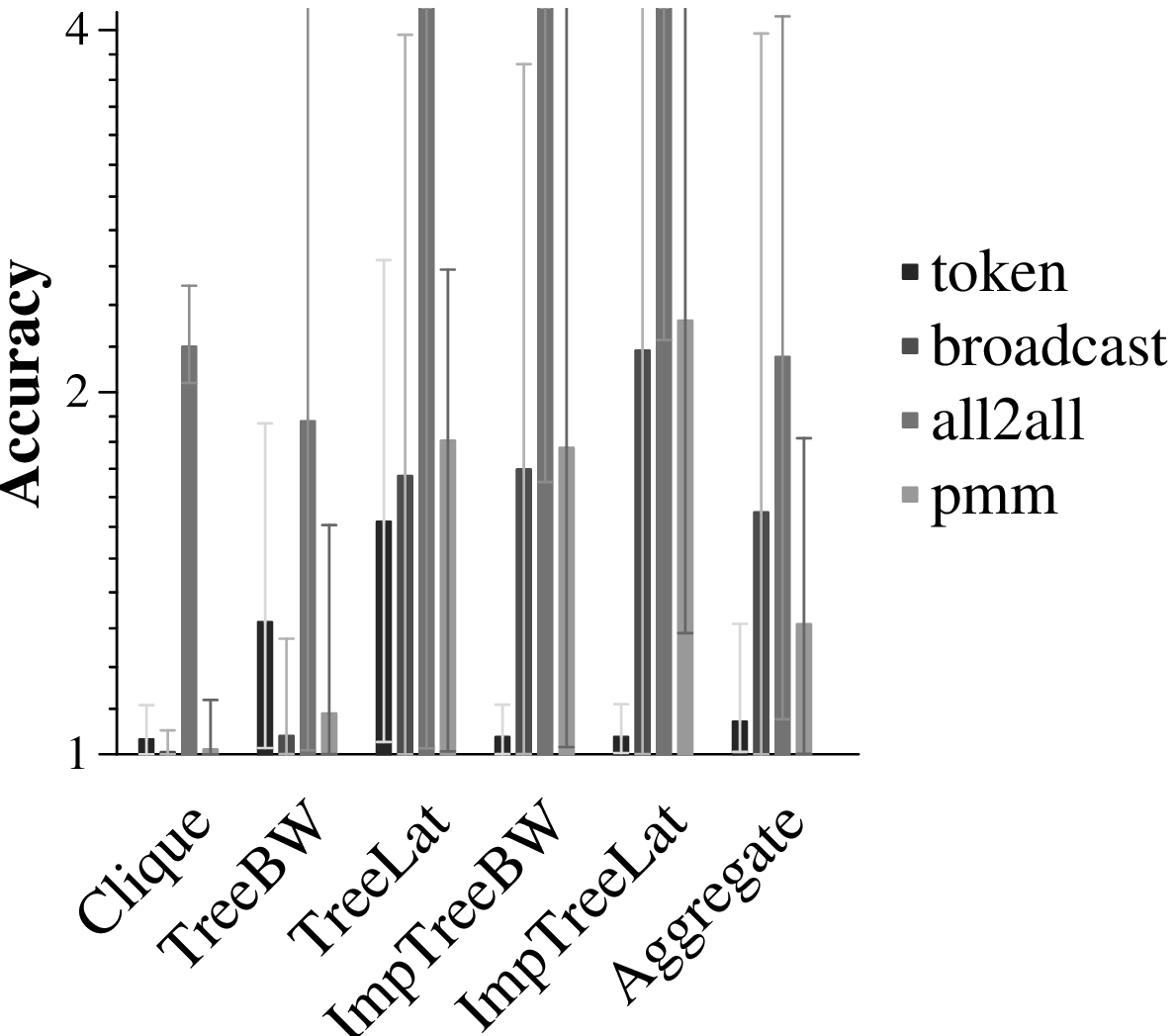}}
\caption{Simulated tests on the GridG platforms, with hidden routers}
\label{fig:gridg_tests_with_routers}
\end{figure*}

The results are shown on Figures~\ref{fig:gridg_tests_without_routers}
and~\ref{fig:gridg_tests_with_routers}. For end-to-end metrics, we
plotted the average accuracy for both latency and bandwidth, and we
also detailed the average accuracy for over- and under-predicted
values. We have also indicated the minimum and maximum values obtained
over all 40 platforms.

Figure~\ref{fig:gridg_tests_without_routers} confirms the results of
the previous section: the improved trees have very good predictive
power, especially \imptreebw, with an average error of 3\% on the most
difficult application, namely \textsc{All2All}. The results of \clique
would be very good too. But as it fails to take interferences into
account, it fails to accurately predict the running time of
\app{all2all}. (Note that the fact that \clique over-estimates the
bandwidth for a few pairs of hosts is due to routing asymmetry in the
original platform.)  We can also see that the basic spanning trees
have better results than in the previous experiment. This is due to
the fact that GridG platforms contain parts that are very tree-like,
which these algorithms are able to reconstruct easily.

However, Figure~\ref{fig:gridg_tests_with_routers} shows that
platforms with hidden routers are much more difficult to reconstruct.
The performance of the clique platform remains the same as before, but
all other algorithms suffer from a severe degradation. It is not clear
yet whether this degradation comes from a wrong view of the topology
of the platform, or from the wrong bandwidth predictions which we can
see on Figure~\ref{fig:gridg_meas_with}.



\section{Conclusion}
\label{sec:conclusion}

In this work, we proposed two new reconstruction algorithms and we
compared them with classical reconstruction algorithms (namely
spanning trees and cliques) through a thorough evaluation
framework. This evaluation framework and the evaluated algorithms are
part of the \alnem project, an application-level measurement and
reconstruction infrastructure, which is freely
available\footnote{\url{http://gforge.inria.fr/plugins/scmcvs/cvsweb.php/contrib/ALNeM/?cvsroot=simgrid}}.

We showed that our \improving procedure, when applied to the maximal
spanning tree on bandwidth, performs very well on instances without
internal routers. The particular efficiency of this algorithm may be
explained by the fact that this is the only algorithm which builds a
non trivial graph (\ie not a clique) while using both the information
on latencies and bandwidths. As a future work, we should design an
algorithm which uses the two types of information simultaneously when
building a model, rather than using one type of information after the
other, as is done to obtain our \imptreebw models.

None of the studied algorithms is fully satisfying in a Grid context,
with hidden internal routing nodes. Our future work is
thus to extend the algorithms to enable them to cope with such a
situation. So far, no algorithm is using any information on
interferences. This should also be addressed as this information
should enable us to design even more efficient network model building
tools.

\bibliographystyle{plain}
\bibliography{alnem}

\end{document}